\documentclass[letterpaper, 10pt,conference]{ieeeconf}

\usepackage{flushend}
\usepackage{epsfig}
\usepackage{graphics}
\usepackage{amsmath}
\usepackage{amssymb}
\usepackage{epsfig,psfrag}
\usepackage{subfigure}
\newtheorem{theorem}{Theorem}
\newtheorem{lemma}{Lemma}
\newtheorem{corollary}{Corollary}

\DeclareMathOperator{\EX}{\mathsf{E}}
\allowdisplaybreaks
\IEEEoverridecommandlockouts
\title{\LARGE \bf Throughput Scaling in Random Wireless Networks:\\ 
A Non-Hierarchical Multipath Routing Strategy}
\author{Awlok Josan, Mingyan Liu, David L. Neuhoff and S. Sandeep Pradhan\\
Electrical Engineering and Computer Science Department\\
University of Michigan, Ann Arbor, MI 48109
\thanks{This work was supported by NSF Grant CCF-0329715.}
}
\begin{document}
\maketitle
\flushend
\begin{abstract}

Franceschetti et al. \cite{franceschetti} have recently shown that per-node throughput in an extended (i.e., geographically expanding), ad hoc wireless network with $\Theta(n)$ randomly distributed nodes and multihop routing can be increased from the $\Omega({1 \over \sqrt{n} \log n})$ scaling demonstrated in the seminal paper of Gupta and Kumar 
\cite{kumar} to $\Omega({1 \over \sqrt{n}})$.  The goal of the present paper is to understand the dependence of this interesting result on the principal new features it introduced relative to Gupta-Kumar:  (1) a capacity-based formula for link transmission bit-rates in terms of received signal-to-interference-and-noise ratio (SINR), instead of the threshold model that positive bit-rate $W$ is attainable when SINR lies above some threshold, and zero bit-rate otherwise;  (2) hierarchical routing from sources to destinations through a system of communal highways, instead of individual direct routes from each source to the corresponding destination; and (3) cell-based routes constructed by percolation rather than by simply interconnecting all cells touched by a straight-line between two end points.  The conclusion of the present paper is that the improved throughput scaling is principally due to the percolation-based routing, which enables shorter hops and, consequently, less interference.  This is established by showing that throughput $\Omega({1 \over \sqrt{n}})$ can be attained by a system that does not employ highways, but instead uses percolation to establish, for each source-destination pair, a set of $\Theta(\log n)$ routes within a narrow routing corridor running from source to destination. As a result, highways are not essential.  In addition, it is shown that throughput  $\Omega({1 \over \sqrt{n}})$ can be attained with the original threshold transmission bit-rate model, provided that node transmission powers are permitted to grow with $n$.  Thus, the benefit of the capacity bit-rate model is simply to permit
the power to remain bounded, even as the network expands.

\end{abstract} 

\section{Introduction}

The problem of asymptotic scalability of throughput in wireless networks has been investigated extensively under different assumptions on the network models. The seminal work of Gupta and Kumar \cite{kumar} demonstrated that per-node throughput $\Omega(1/\sqrt{n\ln n})$ was achievable as the number of nodes in the network, $n$, goes to infinity. 

Franceschetti et al \cite{franceschetti} recently showed that this achievable per-node throughput may be increased.  Specifically, they considered an extended (i.e., geographically expanding) network with approximately $n$ randomly distributed nodes and multihop routing, and demonstrated that achievable per-node throughput can be increased to $\Omega({1 \over \sqrt{n}})$.   

Compared to \cite{kumar}, the construction used in \cite{franceschetti} introduced  several new features.  The first is a capacity-based link transmission rate formula as a function of the received signal-to-interference noise ratio (SINR), instead of the 
threshold-based binary rate model used in \cite{kumar}, where a positive bit-rate $W$ is attainable when the SINR is above some threshold, and zero otherwise.  
(The former requires coding at each hop, while the latter does not.)  The second is a routing hierarchy for data delivery in which data from a source is first delivered (via a single hop) onto a nearby highway -- one of a system of communal highways, each with a horizontal and a vertical segment.  The data is then multihopped along the highway (horizontally then vertically), and finally delivered from the highway to the destination in a single hop.  By contrast, the method used in \cite{kumar} is a simple shortest path type of routing, where a straight line is drawn connecting the source and the destination, and nodes along this line are selected to relay the data, forming an approximately straight line path.  The third difference introduced in \cite{franceschetti} is the use of percolation theory to construct the highways that serve as the main routing fabric in the network.
Indeed, \cite{franceschetti} is the first paper to use percolation theory to establish
network throughput results.
% {\bf (I feel like something more should be said here but not sure what...)}

The primary interest of the present paper is to understand which of the above contribute to the increase in per-node throughput in a fundamental way, i.e., to understand the dependence of this new result on the above new features.   
The conclusion of this paper is that the improved throughput scaling is principally 
due to the percolation-based routing, which enables
shorter hops and, consequently, less interference.  More precisely, the hops along the highways have bounded lengths that do not increase as the network expands.  This would not have been possible if one were to use shortest path routing, the existence of which then invokes a connectivity requirement that would force the hop size to increase as the network expands. 

This conclusion is established by showing that throughput $\Omega({1 \over \sqrt{n}})$ can
be attained by a system that does not employ highways, but rather uses 
percolation to establish, 
for each source-destination (s-d) pair, a set of $\Theta(\log n)$ disjoint routes 
within a narrow routing corridor running from source to destination.
Thus with this multipath routing structure, highways and routing hierarchy are not essential. 
In addition, it is shown that throughput  $\Omega({1 \over \sqrt{n}})$ 
can be attained with the original threshold transmission bit-rate model,
provided the transmission powers of the nodes are permitted to grow with
$n$.  Thus, the benefit of the capacity bit-rate model is simply to permit
the power to remain bounded, even as the network expands.

The remainder of the paper is organized as follows, Section \ref{system} introduces the system and the transmission rate models we use. Section \ref{mainresult} gives our main result and an overview of the proof. The formal proof follows in  sections \ref{percolation}, \ref{throughput}, \ref{load} and \ref{protocol}, which formalize the path construction, data rates, loading factor and the system scheduling, respectively.

\section{System Model}
\label{system}
We consider the random extended network, which consists of a set of nodes distributed over a disk $A_n \subset \mathcal{R}^2$ with radius $\sqrt{n}$, called the network region. We construct the network by placing the nodes according to a Poisson point process of unit intensity over $\mathcal{R}^{2}$ and focusing our attention to the network region $A_n$. We denote the location of the $i$th node by $s_i$. Each node, $s_i$, serves as a source of bits which it wishes to communicate to a destination, denoted by $d_i$, which is chosen randomly from the remaining nodes. Each node may serve as a destination for more than one source. 
Communication is done using a multihop relaying scheme under a slotted time system. There is a transmitter and receiver at each node. All transmitters use the same power $P$, which we get to choose and which may depend upon $n$. We assume that node $j$ receives the transmitted signal from node $i$ with power $P\eta(d_{ij})$, where $\eta$ is a propagation model and $d_{ij}$ is the Euclidean distance between nodes $i$ and $j$. We use the propagation model introduced by Arpacioglu and Haas \cite{haas}, 
\begin{align}
\eta(d) = \frac{1}{(1+d)^\alpha}~,
\end{align}
where $\alpha>2$ is a constant depending upon the channel conditions.

\subsection{Transmission Rate Models}
Let $t$ be a set of simultaneously transmitting nodes. Then the $\mbox{SINR}_{ij}$ (signal to interference and noise ratio) at node $j$ when node $i$ is transmitting to it is given by 
\begin{align*}
\mbox{SINR}_{ij} = \frac{P\eta(d_{ij})}{N_0 + \sum_{\substack{k\in t \\k\neq i}} P\eta(d_{kj})}~.
\end{align*}
We use two different transmission rate models.

\textbf{Model A}
In this model, which was used in \cite{franceschetti}, the transmission rate is equal to the capacity of the wireless channel. That is %This rate model was used by \cite{franceschetti}
the rate (in bits/sec) at which node $i$ can transmit to node $j$ is 
\begin{equation}
R_{ij} = \frac{1}{2} WT \ln(1+\mbox{SINR}_{ij})~,
\end{equation}
where $W$ is the bandwidth and $T$ is length of the time slot.

\textbf{Model B}
In this model, which has been more commonly used in throughput analysis of wireless networks \cite{kumar}--\cite{duarte-melo} % \cite{kumar,haas,duarte-melo}, 
the transmission rate is 
\begin{equation}
R_{ij} = \left\{ \begin{array}{ll}
B & \mbox{if SINR}_{ij} \geq\tau \\
0 & \mbox{else}
\end{array} \right.,
\end{equation}
where $\tau$ is some pre-determined threshold and $B$ is a number less than channel capacity.

\section{Main Result}
\label{mainresult}
%In the following theorem we reprove the result of Franceschetti et al.~\cite{franceschetti} in a non-hierarchical fashion, that is, without the use of highways. We consider both transmission rate models described in Section \ref{system}. 

In the following theorem, which is our main result, we demonstrate the 
achievability of $\Omega(1/\sqrt{n})$ throughput
for both transmission rate models, using a non-hierarchical routing strategy, i.e.,
without the use of highways.

\vspace{1ex}
\begin{theorem}
\label{main}
Under transmission Models A and B, a per-node throughput of 
$\Omega(1/\sqrt{n})$ bits/sec
is achievable in the random extended network.
Under Model A the throughput is achievable with any constant finite power $P$ at each node, whereas under Model B the throughput is achievable only if power $P$ increases to infinity as $n \rightarrow \infty$.
\end{theorem}

\vspace{1ex}
We now give an overview of the proof, details of which are in subsequent sections. For each s-d pair we find with high probability $\Omega(\ln n)$ disjoint routes (i.e., a sequence of hops from node to node) from source to destination such that 

\vspace{1 ex}

{\leftskip = .25in
\noindent 1.   each route consists of a \emph{draining hop} from the source,
a \emph{path} consisting of a sequence of intermediate hops, and a
\emph{delivery hop} ending at the destination,  \newline
\noindent 2.  the first hop, i.e., the draining hop, has length   $O(\ln n)$ and
extends from the source to the first node of the path, \newline
\noindent 3.  the last hop, i.e. the delivery hop, has length $O(\ln n)$, and
extends from the last node of the path to the destination. \newline
\noindent 4.  all intermediate hops have lengths bounded by a constant 
not depending on $n$.
\par }

%\vspace{-1.5ex}

%\vspace{1 ex}

%{\leftskip = .25in
%%
%\noindent 1.   % the length of the first hop from the source is $O(\ln n)$,\newline
%the first hop from the source, which is called a \emph{draining hop}, has length 
% $O(\ln n)$,\newline
%%
%2.   % the length of the last hop that delivers the packet to the destination is $O(\ln n)$ and \newline
%the last hop before the destination, which is called a \emph{delivery hop}, has length 
% $O(\ln n)$,\newline
%%
%3. all \emph{intermediate hops}, from the second node to the next-to-last node, have  lengths bounded by a constant not depending on $n$.  This sequence
%of bounded length hops will be called a \emph{path}.  %\newline 
%%\indent 3. all intermediate hop lengths are bounded by a constant not depending on $n$. \newline
%%
%Thus, each route consists of a draining hop, a path and a delivery hop. \newline
%\par }

\vspace{1ex}
\noindent 
To make the analysis tractable, we modify these paths slightly in a way that
preserves their distance properties, but does not necessarily preserve their
disjointness.
We then show that for each s-d pair, a rate of $\Omega(1/(\sqrt{n}\ln n))$ is sustainable on each hop of each of its modified paths.  To do this, we show that the maximum number of source-destination paths on which an intermediate node can lie is $O(\sqrt{n}\ln n)$.  From Item 4 above, the intermediate nodes, with the exception of the delivery node, transmit over a bounded distance.  Theorem 3 of \cite{franceschetti} 
%%% formerly Thm 4
showed that when transmitting over a bounded distance, nodes can maintain a throughput of $\Omega(1)$. Thus for each s-d pair an intermediate node can sustain a throughput of $\Omega(1)\times 1/O(\sqrt{n} \ln n) = \Omega(1/(\sqrt{n} \ln n))$.

Next, using Theorem 3 of \cite{franceschetti} 
%%% formerly Thm 4
again, we show that a source can transmit data at rate $\Omega(1/\sqrt{n})$ in a way that will be received by a node on each of the $\Omega(\ln n )$ paths for the s-d pair. Through this node, each path then takes a share of this rate equal to  $\Omega(1/(\sqrt{n}\ln n))$. Therefore, the source is able to drain onto each of the $\Omega(\ln n)$ paths at rate $\Omega(1/(\sqrt{n}\ln n))$. Similarly, delivery nodes can deliver data to the destination at a rate of $\Omega(1/(\sqrt{n}\ln n))$ from each path.
%Next we show that a source can transmit data onto all of these paths at a rate of $\Omega(1 /\sqrt{n})$. Since there are $\Omega(\log n)$ paths, a source can thus transmit data onto every single path at a rate $\Omega(1/(\sqrt{n}\ln n)$. Similarly, on the delivery side, a node can deliver to the destination at rate $\Omega(1/\sqrt{n})$. Setting up a TDMA schedule for all paths of a source-destination pair, we can thus achieve a delivery rate of $\Omega(1/(\sqrt{n} \ln n))$ per path.  

Combining the above results we see that, for each source-destination pair we have $\Omega(\ln n)$ routes, each of which can sustain a rate of $\Omega(1/(\sqrt{n}\ln n))$. Thus the per-node throughput is given by $\Omega(\ln n)\times \Omega(1/(\sqrt{n}\ln n)) = \Omega(1/\sqrt{n})$.

\section{Path Construction via Percolation}
\label{percolation}

In this section we show that, with probability approaching 1 as $n \rightarrow \infty$, there exist $\Omega(\ln n)$ suitable disjoint paths for each source-destination pair. Here the probability is %over the node locations and the set of all possible source-destination pair sets. 
with respect to the Poisson point process for node locations and the random
destination assigned to each source node.
To do this, we use the percolation approach that was used in \cite{franceschetti}
to establish the existence of suitable highways.  Here we apply
approach to find a set of suitable paths for each source-destination pair. 

%To do this, we require a well known result from percolation theory. The approach here is similar to the one used in \cite{franceschetti} to establish the existence of
%suitable highways. 

Since we need to show the existence of paths for every s-d pair, we first need to upper bound the number of nodes in the network region $A_n$, which we denote $N_n$. 

\vspace{1ex}
\begin{lemma}
\label{lem-numnodes}
The probability that the number of nodes, $N_n$, in the network region $A_n$ is less than $2\pi n$ goes to 1 as $n$ goes to infinity.
\end{lemma}
\vspace{0.5em}

\noindent \emph{Proof:}
The number of nodes in the network region, $N_n$, is a Poisson random variable with mean $\pi n$. Applying the Chernoff bound gives,
\begin{align*}
\Pr(N_n > 2\pi n) &\leq e^{-2s\pi n} \EX[e^{sN_n}] \\
&= e^{-2s\pi n} e^{\pi n(e^s-1)}
\end{align*}
for all $s>0$. Choosing $s=1$ gives
\begin{align*}
\Pr(N_n \leq 2\pi n) &\geq 1 -  e^{-2\pi n}e^{\pi n(e-1)} \\
& = 1 - e^{\pi n(3-e)} \\ 
&\rightarrow 1 \mbox{  as  } n \rightarrow \infty ~. 
\end{align*}

\vspace{-4.3ex} \hfill $\square$

\begin{figure}
\centering
\subfigure[Tessellation of a rectangular routing corridor with diamonds of side length $c$.]
{
\includegraphics[width=3in]{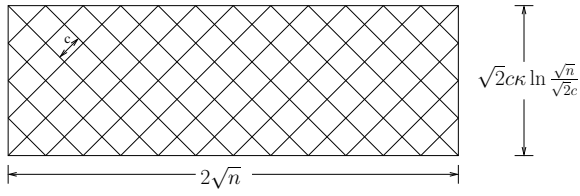}
\label{fig-slab}
%\caption{Tessalation of a rectangular slab with sqaures of side $c$.}
}
%\vspace{1cm}
\subfigure[Paths crossing the routing corridor from left to right are composed from horizontal and vertical edges, shown as dashed lines.]
{
\includegraphics[width=3in]{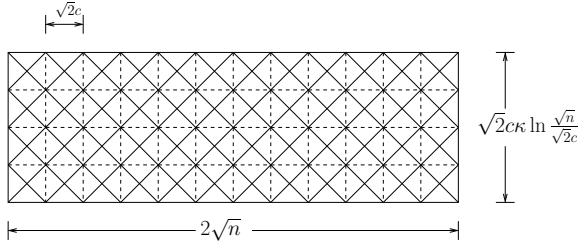}
%\caption{Tessalation of a rectangular slab with sqaures of side $c$.}
\label{fig-slab1}
}
\caption{Routing corridor setup for finding paths for a given s-d pair.}
%\label{fig-slab}
\end{figure}

\vspace{1.5ex} 
Next we prove that for a given s-d pair, there are $\Omega(\ln n)$ disjoint paths such that the distance to (from) each path from (to) the source (destination) is $O(\ln n)$, and that every intermediate hop along each path is of length $O(1)$, i.e. its length is upper bounded by a constant independent of $n$.
To show this, we consider a rectangular \emph{routing corridor} of dimensions $2\sqrt{n}\times \sqrt{2}c\kappa\ln\frac{\sqrt{n}}{\sqrt{2}c}$ in $\mathcal{R}^2$ that includes both s and d, where $c,\kappa>0$ are constants to be chosen later. 

Tessellate this routing corridor with diamonds of side $c$ as shown in Figure \ref{fig-slab}. Then for any given diamond, 
\begin{align*}
\Pr(\mbox{diamond contains at least one node}) = 1 - e^{-c^2} \triangleq p ~.
\end{align*}
If a diamond contains at least one node, it is said to be \textit{open},  and \textit{closed} otherwise. Draw horizontal edges across half the diamonds and vertical edges across the others in the manner shown in Figure \ref{fig-slab1}. An edge is considered \textit{open} if it lies in an open diamond, and \textit{closed} otherwise.  Define a path as a sequence of connected edges, horizontal or vertical.  A  path is said to be open if it contains only open edges. We will show that there are $\Omega(\ln n)$ disjoint open paths crossing the routing corridor lengthwise, i.e. beginning at the left and ending at the right side of the routing corridor.

Let $I_m$ be the event that there exist at least $m$ disjoint open paths that cross the routing corridor lengthwise.

The following lemma, whose proof can be found in the proof of Theorem 5 of \cite{franceschetti}
%formerly Thm 3
 is based on an important result from percolation theory. 

\vspace{1ex}
\begin{lemma}
\label{lem-interm}
Given arbitrary constants $\kappa,c >0$, there exists a strictly positive constant $\beta =\beta(c,\kappa)$ such that 
\begin{equation}
\Pr(I_{m}) \geq 1 - \frac{4}{3}\left(\frac{n}{2c^2}\right)^a
\end{equation}
where $m=\beta \kappa \ln\frac{\sqrt{n}}{\sqrt{2}c}$ and $a = \frac{1}{2}\left((\beta - 1) \kappa c^2+\kappa \ln 6 +1\right)$.
\end{lemma}
%\vspace{0.5em}
%\noindent \emph{Proof:}
%This lemma is proved in the proof of Theorem 5 in \cite{franceschetti}.
%%%% formerly Thm 3
%\hfill $\square$

\vspace{1ex}
We now set up a routing corridor for each s-d pair. The following theorem demonstrates that when $n$ is large, with high probability there are $\Omega(\ln n)$ disjoint paths in each one of those corridors.
\vspace{1ex}
\begin{theorem}
\label{thm-paths}
Given $\kappa > 0$ and $c > \ln 6 + 4/\kappa$, there exists a strictly positive constant $\beta(c,\kappa)$ such that if for every $n$ we are given at most $\lceil2\pi n\rceil$ 
routing corridors of dimensions $2\sqrt{n}\times \sqrt{2}c\kappa\ln\frac{\sqrt{n}}{\sqrt{2}c}$ in $\mathcal{R}^2$, then with probability approaching one there exist $m=\beta \kappa \ln \frac{\sqrt{n}}{\sqrt{2}c}$ disjoint open lengthwise crossing paths within each of the routing corridors. 
\end{theorem}

\vspace{1ex}
Observe that when $n$ are large, the routing corridors are quite narrow.

\vspace{1ex}
\noindent \emph{Proof:}
We prove this theorem using Lemma \ref{lem-interm} and the union bound. It suffices to assume that we have $\lceil 2\pi n\rceil$ routing corridors. Then
\begin{align*}
\Pr(\mbox{all }\lceil2\pi n\rceil &\mbox{ routing corridors have $m$ disjoint open paths}) \\
&= 1 - \Pr(\mbox{at least one routing corridor has} \\
&\qquad\qquad\qquad\mbox{less that $m$ disjoint open paths}) \\ 
&\geq 1 - \sum_{i=1}^{\lceil 2\pi n\rceil} \Pr(\mbox{ith routing corridor has}\\
&\qquad\qquad\qquad\mbox{less than $m$ disjoint open paths}) \\
&\geq 1 - \lceil 2\pi n \rceil \cdot \Pr(\mbox{a routing corridor pair has}\\
&\qquad\qquad\qquad\mbox{less than $m$ disjoint open paths}) \\
&= 1- \lceil2\pi n \rceil(1-\Pr(I_{m}))\\
&\geq 1-  8n \cdot \frac{4}{3}\left(\frac{n}{2c^2}\right)^a \\
&= 1 - \frac{32}{3(2c^2)^a} n^{a+1}
\end{align*}
where the first inequality follows from the union-bound and the second inequality uses Lemma \ref{lem-interm}. Note that the above expression goes to one as $n$ tends to infinity if $a<-1$. Given $\kappa > 0$ and $c> \ln 6 +4/\kappa$, choosing $\beta(c,\kappa) = 1-(\kappa \ln 6 + 4)/(\kappa c^2) >0$ results in $a<-1$.
\hfill $\square$

\begin{figure}
\centering
\includegraphics[width=3in]{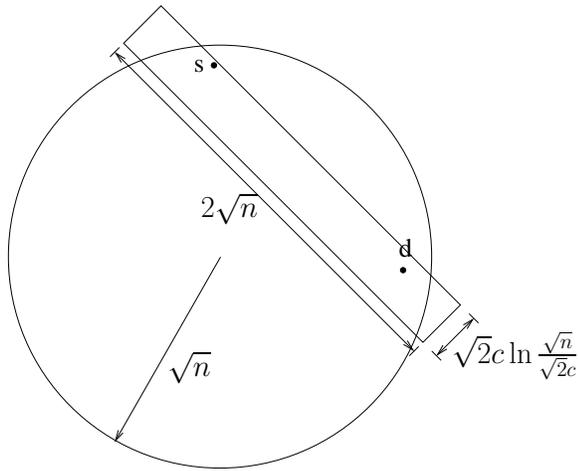}
\caption{For a given s-d pair the orientation of the routing corridor on the network region.}
\label{fig-network}
\end{figure}

\vspace{1ex}
\begin{corollary}
Given $\kappa > 0$ and $c > \ln 6 + 4/\kappa$, there exists a strictly positive constant $\beta(c,\kappa)>0$ such that with probability approaching one there exist $\Omega(\ln n)$ disjoint open paths for each s-d pair such that the distance of any path from the source and destination is less than $\sqrt{2}c \kappa \ln (\sqrt{n}/\sqrt{2}c)$ and every intermediate hop has length less than $\sqrt{5}c$.
\end{corollary}

\vspace{0.5em}
\noindent \emph{Proof:}
For any given s-d pair, consider a routing corridor with the aforementioned dimensions such that it contains both source and destination and that the portion of the routing corridor that intersects the network region is as high as possible (see Figure \ref{fig-network}). According to Lemma \ref{lem-interm},  with high probability there are $\Omega(\ln n)$ disjoint open paths that cross the routing corridor lengthwise. Now consider the part of the routing corridor that lies within the network region. 
Since there are $\Omega(\ln n)$ disjoint open paths that cross the routing corridor lengthwise, there will be $\Omega(\ln n)$ disjoint open paths in the truncated region as well. Also, since the width of the routing corridor is $\sqrt{2}c\kappa \ln\frac{\sqrt{n}}{\sqrt{2}c}$ , the minimum distances of each of these paths from the source and the destination is less than $\sqrt{2}c\kappa \ln\frac{\sqrt{n}}{\sqrt{2}c}$.  Also, using a geometric argument, it is easy to see that any intermediate hop has less $\sqrt{5}c$ or less.

Theorem \ref{thm-paths} shows the existence of paths for a number of routing corridors no larger than $\lceil2\pi n\rceil$. Using the above construction for every s-d pair and combining with the fact that the number of s-d pairs is less the $2\pi n$ with high probability (Lemma \ref{lem-numnodes}) completes the proof of the corollary.
\mbox{ } \mbox{ }  \hfill $\square$

\vspace{.5ex}

As suggested earlier, for tractability we need to modify the paths provided by
the corollary.   Ignoring the previous tesselations of routing corridors, consider now a tessellation of the entire network region into squares of side $c$.   
If a square has multiple nodes in it, we designate one node as the \emph{relay node}.  
Now, for every hop of every s-d path, if the node that is to transmit is not the designated relay node for the square, we replace it with the designated relay node. 
In this way we obtain a set of $\Omega(\ln n)$ paths for each s-d pair such that each
source (destination) is within $O(\ln n)$ of each of its paths.   Note, however, that now  the maximum intermediate hop length has been increased to $(\sqrt{5}+\sqrt{2})c$. 
Moreover, the paths corresponding to one s-d pair might no longer be disjoint. 
For example, in two originally disjoint paths there might be a node
in one path and a node in the other that are contained in adjacent diamonds in the
original tesselation of the routing corridor, but are in the same square of
the new tesselation of the entire network region.  In this case, the two modified
paths share a common relay node.  

\section{Data Rates}
\label{throughput}

We begin this section by finding a lower bound on the per-node transfer rate when for some $D>0$ every node has to send data to all nodes within distance $D$ of itself. This involves setting up a TDMA schedule so as to limit the number of simultaneous transmissions taking place, which in turn limits the interference.  Corollaries are then given for use in the proof of the Theorem \ref{main}.

For transmission rate Model A, Theorem 3 of \cite{franceschetti} %%% was Thm 4
can be used.  The following extends this theorem to transmission rate Model B. 

\vspace{1ex}
\begin{theorem}
\label{thm-throughput}
Given $c>0$, given a tessellation of the net-work into squares with sides of length $c$, and given an integer $d>0$ there exists a rate $R(d)=\Omega(d^{-\alpha-2})$ using Model A and $R(d)=\Omega(d^{-2})$ using Model B such that one node in each square can
% 
%transmit data at rate $R(d)$ that is successfully received by 
%transfer data at rate $R(d)$ that is successfully received by 
successfully transfer data at rate $R(d)$ to 
any node located in any square within Manhattan distance $d$ of the originating square (i.e. $d$ or fewer horizontal and/or vertical steps).

The asymptotic behavior of the rate under Model A can be attained by any fixed finite power at each node. However to achieve the rate under Model B we have to let power $P$ go to infinity as $d$ tends to infinity.
\end{theorem}

\vspace{1ex}

\noindent \emph{Proof:}
For Model A the proof is given in \cite[Theorem 3]{franceschetti}, %%% was Thm 4
and for the extension to Model B, we now make a similar construction. We consider a partition of the network region into super-squares, each composed of $k^2$ smaller squares, for
some $k$ to be chosen later. We index the squares in each super-square starting in the lower left corner, moving horizontally in the bottom row from left to right, and then in the row above it from left to right, and so on. We set up a TDMA schedule of $k^2$ slots such that in the $i$th slot, from every square indexed by $i$, precisely one node can transmit. 

\begin{figure}
\centering
\includegraphics[width=2.5in]{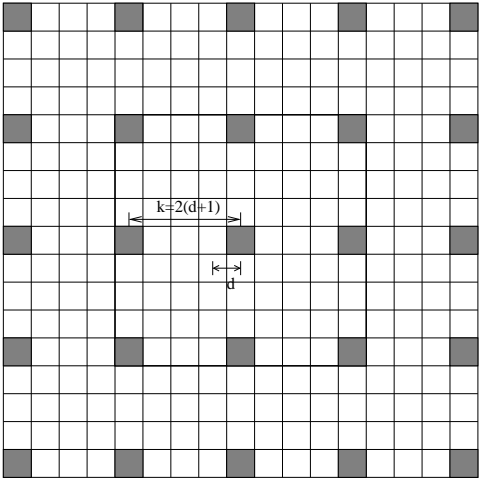}
\caption{Construction for lower bound on SINR. The shaded square at the center is the actual signal, all other shaded squares are interfering transmitters. In the above figure $d=1$.}
\label{fig-throughput}
\end{figure}

Consider a transmitter-receiver pair separated by $d$ squares. Choosing $k=x(d+1)$, where 
\begin{align*}
x=\max(2,\lceil (16\tau \gamma)^{1/\alpha} (1+1/(2c)) \rceil) ~,
\end{align*} 
and $\gamma = \sum_{i=1}^{\infty}(i-1/2)^{-\alpha+1}$, we can see that the closest 8 interferers are at least $x(d+1)-d$ squares away, the next closest 16 interferers are at least $2x(d+1)-d$ squares away, and so on (Figure \ref{fig-throughput}). The power from interfering nodes can thus be upper bounded as 
\begin{align*}
P_I(d) &\leq \sum_{i=1}^{\infty} 8iP\eta(c(ix(d+1)-d)) \\
& = \sum_{i=1}^\infty \frac{8iP}{(1+c(ix(d+1)-d))^\alpha} \\
& \leq 8P \sum_{i=1}^{\infty} \frac{i}{(c(d+1)(ix-1))^{\alpha}} \\
& = \frac{8P}{c^{\alpha}(d+1)^{\alpha}x^{\alpha}} \sum_{i=1}^{\infty} \frac{i}{(i-1/x)^{-\alpha}} \\
& \leq \frac{8P}{c^{\alpha}(d+1)^{\alpha}x^{\alpha}} \sum_{i=1}^{\infty} \frac{2(i-1/2)}{(i-1/2)^{-\alpha}} \\
&= \frac{16P}{c^{\alpha}(d+1)^{\alpha}x^{\alpha}} \sum_{i=1}^{\infty}(i-1/2)^{-\alpha+1}  \\
& = \frac{16P\gamma}{c^\alpha(d+1)^{\alpha}x^{\alpha}} ~.
%&\leq \frac{16P\gamma}{(2c)^\alpha x^{\alpha}} ~,
%&\leq \sum_{i=1}^{\infty} \frac{16iPc^{-\alpha}(2i-1)^{-\alpha}(d+1)^{-\alpha} \\
%& = P[c(d+1)]^{-\alpha}K(d)
\end{align*}
%where the last inequality comes from the fact that $d\geq 1$~.

Next we lower bound the signal power at the receiver. The Euclidean distance between the transmitter and receiver is at most $c(d+1)$. Thus the signal power, $P_S(d)$, satisfies
\begin{align*}
P_S(d) &\geq P \eta(c(d+1)) \\
& = \frac{P}{(1+c(d+1))^\alpha} ~.
\end{align*}

Using the above two bounds we obtain a bound on the SINR:
\begin{align*}
\mbox{SINR}(d) &= \frac{P_S(d)}{N_0+P_I(d)} \\
&\geq \frac{P(1+c(d+1))^{-\alpha}}{N_0+16P\gamma (2c)^{-\alpha}x^{-\alpha}} \\
%\mbox{SINR}(d) 
&= \left(\left(1+\frac{1}{c(d+1)}\right)^{\alpha}\frac{N_0}{P} \right.\\
& \left.\qquad\qquad\qquad\quad+ \left(1+\frac{1}{c(d+1)}\right)^{\alpha}\frac{16\gamma}{x^\alpha}\right)^{-1}  ~. \\
%&\geq \left(\left(1+\frac{1}{2c}\right)^{\alpha}\frac{N_0}{P} + \left(1+\frac{1}{2c}\right)^{\alpha}\frac{16\gamma}{x^\alpha}\right)^{-1}  ~, \\
\end{align*}
%As $d$ grows to infinity, $K(d)$ goes to zero. Also if $P$ grows at least as fast as $(d+1)^{\alpha}$, then it can be chosen so as make SINR greater than any predetermined threshold $\tau$. In this case according to Model B, one node in each square can transmit at rate 1 in such a way that all nodes within Manhattan distance $d$ will successfully receive the transmissions. Since each sqaure is allowed to have a transmitting node once every $k^2=4(d+1)^2$ time slots, to get the asymptotic behavior we need to divide the above transfer rate by $d^2$. Thus under Model B, $R(d) = \Omega(d^{-2})$ is attainable.
%where the last inequality is true due to the fact that $d\geq1$. 
It can be easily shown that the second term in the above equation is less than $1/\tau$. Choosing $P$ large enough that the sum of two terms still remains less than $1/\tau$ results in SINR$>\tau$.
In this case according to Model B, one node in each square can transmit at rate 1 in such a way that all nodes within Manhattan distance $d$ will successfully receive the transmissions. Since each square is allowed to have a transmitting node once every $k^2=x^2(d+1)^2$ time slots, to get the asymptotic behavior we need to divide the above transfer rate by $d^2$. Thus under Model B, $R(d) = \Omega(d^{-2})$ is attainable.
\hfill $\square$

\vspace{1ex} We now give a corollary to the above theorem that will be used to show an achievable data delivery rate to the destination.

\vspace{1ex}
\begin{corollary}
\label{cor-recrate}
Given $c>0$, given a tessellation of the network into squares with sides of length $c$, and given an integer $d>0$ there exists a rate $R(d)=\Omega(d^{-\alpha-2})$ for Model A and $R(d)=\Omega(d^{-2})$ for Model B such that one node in each square can receive data at rate $R(d)$ from a transmitter located in any square within Manhattan distance  $d$ of the receiving square (i.e. $d$ or fewer horizontal and/or vertical steps).
%Given an integer $d>0$ there exists a TDMA schedule which allows one node in each square to receive from a transmitter located within a radius of $d$ squares at a rate $R(d)$ which is independent of $n$. The asymptotic behavior of $R(d)$ as $d$ goes to infinity is give by $R(d) = \Omega(d^{-\alpha-2})$ under Model A and $R(d) = \Omega(d^{-2})$ under Model B.
\end{corollary}

\vspace{1ex}
\noindent \emph{Proof}
The proof is obtained by switching the role of transmitters and receivers in the proof of the previous theorem.
\mbox{ } \mbox{ }  \hfill $\square$

We conclude this section with three corollaries that use Theorem \ref{thm-throughput} 
to establish rates at which, respectively, draining, delivery and transmission along the
intermediate hops can proceed.

\vspace{1ex}
\begin{corollary}
\label{cor-drain}
With probability approaching one, every source node in the network can transmit to every one of the $\Omega(\ln n)$ paths in its corresponding routing corridor at a rate 
$\Omega( (\ln n)^{-\alpha-4})$ under transmission Model A, and $\Omega((\ln n)^{-4})$ under Model B.
\end{corollary}

\vspace{1ex}
\noindent \emph{Proof:}
First, for Model A, consider the tessellation of $A_n$ into squares of side length $c$. Consider also any one source node. Since the Manhattan distance from this source to each of its paths is less than $\phi \ln n$, for some $\phi > 0$,  if this node is the only node within its square then Theorem~\ref{thm-throughput} with $d=\phi \ln n$ implies it can transmit data that is successfully received by a node on each of its paths at rate
\begin{align*}
R(\phi \ln n) = \Omega( (\ln n)^{-\alpha-2}) ~.
\end{align*}
It is therefore decided that nodes will transmit at rate $\Theta((\ln n)^{-\alpha-3})$, and since each path takes responsibility for relaying an equal share of this data, each path is responsible to relay $\Theta((\ln n)^{-\alpha-4})$.
When $n$ is large, with high probability the number of nodes in a square of size $c$ is $O(\ln n)$ \cite[Lemma 1]{franceschetti}.  %%% was Lemma 3
Every node can actually transmit data at  rate of $\Theta((\ln n)^{-\alpha-4})$. The proof for Model B follows similar arguments.
\hfill $\square$

\vspace{1ex}
\begin{corollary}
\label{cor-deliver}
With probability approaching one, every destination node in the network can receive data from every one of the $\Omega(\ln n)$ paths in its corresponding routing corridor at a rate $\Omega(\ln n)^{-\alpha-5})$ under Model A, and $\Omega((\ln n)^{-5})$ under Model B.
\end{corollary}

\vspace{1ex}
\noindent \emph{Proof:}
First, for Model A, consider a tessellation of $A_n$ into squares of side length $c$. Consider any one destination node and one of the source nodes that corresponds to that destination. Since the distance to the destination from each of its paths is less than $\phi \ln n$, for some $\phi>0$, if this node is the only node within its square then Corollary \ref{cor-drain} implies that data can be successfully received by the destination at rate $R(\phi \ln n) = \Omega((\ln n)^{-\alpha-2})$. It is therefore decided that nodes delivering data to this destination will transmit at rate $\Theta((\ln n)^{-\alpha-2})$. Using the Chernoff bound we can easily see that the number of sources that choose any given node as its destination is $O(\ln n)$ with high probability. Setting up a TDMA scheme in which each epoch consisting of $O((\ln n)^2)$ slots would allow the destination to receive from every path of every source that selects the given node as its destination at least once in every epoch. Thus a destination can receive at rate $\Omega((\ln n)^{\-\alpha-4})$. 
When $n$ is large with high probability the number of nodes in a square of size $c$ is $O(\ln n)$ \cite[Lemma 1]{franceschetti}. %%% was Lemma 3
Thus every node can receive data at rate $\Omega((\ln n)^{-\alpha-5})$.
%Substituting $d=\phi \ln n$ in Corollary \ref{cor-recrate} gives $R(\phi \ln n) = \Omega((\ln n)^{-\alpha-2})$. Setting up a TDMA schedule for all $\ln n$ paths gives a receiving rate of $\Omega((\ln n)^{-\alpha-3})$ per path. Once again the number of destinations in the square is $O(\ln n)$, thus every destination can receive at a rate of $\Omega((\ln n)^{-\alpha-4})$. 
The proof for Model B follows similar arguments.
\hfill $\square$

\vspace{1ex}
\begin{corollary}
\label{cor-relay}
Given $c>0$, and a tessellation of $A_n$ into squares of side length $c$, one node in every square can transmit to every node located within distance $O(1)$, i.e., distance is upper bounded by a constant that does not depend upon $n$, at a constant rate that does not depend upon $n$.
\end{corollary}

\vspace{1ex}
\noindent \emph{Proof:}
First consider Model A. From Theorem \ref{thm-throughput} we know that one node in every square can achieve a rate of $\Omega(d^{-\alpha-2})$ while transmitting to every node located within Manhattan distance $d$ of the originating square. For transmissions over distance that is upper bounded by a constant not depending upon $n$, $d$ would be a constant. Hence rate $\Omega(1)$ is achievable over constant distance. The proof for Model B follows similar arguments.
\hfill $\square$  

\section{Loading Factor}
\label{load}

The \emph{loading factor} of a designated relay node is the number
of s-d paths on which it lies.  We also consider it to be the loading factor
of the square containing the relay node.  
In this section we find a probabilistic upper bound to the maximum loading factor
among all squares, which then upper bounds the maximum loading factor
of all relay nodes.  

%We show that the loading factor of any square in the network is $O(\sqrt{n}\ln n)$.

Let $L_i(n)$ represent the loading factor of the $i$th square, and let $L(n) = \max_i L_i(n)$.   
%
%To find an upper bound 
%
We observe that if an s-d pair contributes a path or paths to the $L_i(n)$, then
it must be that the corresponding routing corridor intersects the $i$th square. 
%Even then, it will contribute a path to $L_i$ only if there is a tentative path
Now, we observe that if the $i$th square intersects a given s-d routing corridor, it can, at most, intersect 9 diamonds of the routing corridor tessellation. Recall that the tentative paths for a given s-d pair are disjoint, i.e. a diamond of the s-d routing corridor can lie on only one tentative path. Thus, if the $i$th square intersects the s-d routing corridor it may have to service at most 9 paths corresponding to that s-d pair. 
%Moreover, one or more of the diamonds tessellating the routing corridor must intersect the square. 
%
%because tentative paths for any one s-d pair are disjoint, each diamond of the routing corridor tessellation can lie on at most one.
%
%however because the square network tessellation is unsynchronized more than one
%tentative paths can contribute nodes to the ith square.  
%
%each square of network tessellation can intersect at most 9 diamonds
%
%therefore each s-d pair whose routing corridor intersects the $i$th square can contribute
%at most 9 paths to $L_i(n)$.   

%For every s-d pair we check if the square intersects the routing corridor corresponding to that s-d. If it does, it can at most intersect 9 diamonds of the routing corridor tessellation, and hence nodes in the square can lie on at most 9 paths for that s-d pair. 
Therefore as an upper bound to $L(n)$, we upper bound the number of s-d routing corridors that intersect any given square and multiply that number by 9.

\vspace{1ex}
\begin{theorem}
\label{thm-maxpath}
For a tessellation of the network region into squares of side $c$, there exists a constant $\delta$ such that
\begin{equation*}
\Pr(L(n) \leq \delta \sqrt{n}\ln n) \rightarrow 1 \textrm{  as } n \rightarrow \infty ~.
\end{equation*} 
\end{theorem}

\vspace{1.5ex}
\noindent \emph{Proof:}
\begin{align}
\Pr(L(n) \leq \delta \sqrt{n} \ln n) &= \Pr(\max_i L_i(n) \leq \delta \sqrt{n} \ln n) \nonumber \\
&\geq 1 - \sum_{i=1}^{M_n} \Pr(L_i(n) > \delta \sqrt{n} \ln n)
\label{eq-maxpath}
\end{align}
where $M_n \approx \frac{\pi n}{c^2}$ is the number of squares in the network region.
We have $L_i \leq 9\sum_{j=1}^{N_n} A_{ij}$ where $A_{ij}=1$ if the $i$th square intersects the routing corridor corresponding to the $j$th s-d pair and $A_{ij} = 0$ otherwise. Note that for a given $i$, $A_{i1},A_{i2}...$ are independent and identically distributed.
However the $L_i$'s are not identically distributed. Instead $L_i$ will generally have a higher value for squares near the center of $A_n$ than its boundary. 
%However,using Lemma \ref{lem-maxprob}, we can upper bound the probability $\Pr(A_{ij}) \triangleq p_{n,i} \leq p_n$.
The following lemma, which gives a uniform upper bound to $p_{n,i}\triangleq \Pr(A_{ij}=1)$, will be used to find a lower bound to the term $\Pr(L_i > \delta \sqrt{n} \ln n)$ that appears in (\ref{eq-maxpath}).

\begin{figure}
\centering
\includegraphics[width=3in]{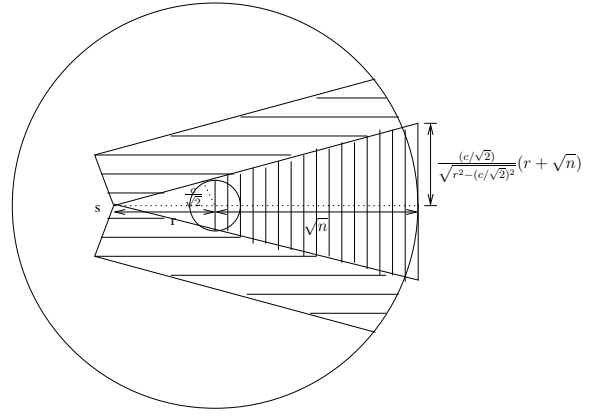}
\caption{The $i$th square lies on a s-d path only if the destination lies in the striped region.}
\label{fig-maxprob}
\end{figure}

\vspace{1ex}
\begin{lemma} 
\label{lem-maxprob}
Given $c>1/\sqrt{2}$ there exists $\mu$ such that 
\begin{equation}
    p_{n,i} \leq p_n \triangleq \mu \ln n/\sqrt{n} ~, \mbox{ for all } n,i ~.
\end{equation}
\end{lemma}

\vspace{1ex}
\noindent \emph{Proof:}
We setup a polar coordinate system such that the origin lies at the center of the network region.  
As the probability of intersection of a square by a random s-d pair routing corridor is highest at the center, we consider the $i$th square to lie at the center of the network region, i.e., to contain the origin.  Since such a square of side $c$ is completely contained in a circle of radius $c/\sqrt{2}$, we upper bound $p_{n,i}$ by the probability of a random s-d routing corridor intersecting a circle of radius $c/\sqrt{2}$ centered at the origin. 

For a source located at $(r,\theta)$, the probability that square $i$ is intersected by the s-d pair routing corridor is upper bounded by the probability of the destination lying in the striped regions of Figure \ref{fig-maxprob}. Since the diameter of the network region is $2\sqrt{n}$, the area of the horizontally striped regions can be upper bounded by $2\cdot2\sqrt{n}\cdot\sqrt{2}c\kappa\ln\frac{\sqrt{n}}{\sqrt{2}c}$. Since $c>1/\sqrt{2}$ the upper bound can be relaxed to $2\sqrt{n}\cdot\sqrt{2}c\kappa\ln n$. Also, the area of the vertically striped portion is $\frac{(c/\sqrt{2})}{\sqrt{r^2-(c/\sqrt{2})^2}}(r+\sqrt{n})^2$. Therefore
\begin{align*}
&\Pr\left(\mbox{s-d routing corridor intersects square i}|s=(r,\theta)\right) \\
&\quad\leq 
\left\{ \begin{array}{ll}
1 & \mbox{ if $r\leq c$} \\
\frac{2\sqrt{2}\kappa c}{\pi} \frac{\ln n}{\sqrt{n}} + \frac{(c/\sqrt{2})}{\sqrt{r^2-(c/\sqrt{2})^2}}\frac{(r+\sqrt{n})^2}{\pi n} & \mbox{  otherwise}\\
%& = \mu \frac{\ln n}{\sqrt{n}}
\end{array} \right.
\end{align*}
%\enlargethispage{-1.6in}
Since the joint probability density of the polar coordinate locations is $p(r,\theta) = \frac{2r}{n}\frac{1}{2\pi}$, we have 
\begin{align*}
  p_{n,i} &=  \int_0^{2\pi}  \!\! \int_0^{\sqrt{n}} 
          \!\!\!  \Pr \! \Big( \!\!  \begin{array}{c}
                   \mbox{\small s-d routing corridor} \\  \mbox{\small intersects square } i   
                   \end{array} 
                    \!\! \Big|  s=(r,\theta) \Big)   p(r,\theta) \, dr  d\theta
                    \nonumber \\
%p_{n,i} &= \int_0^{2\pi}\int_0^{\sqrt{n}} Pr(\mbox{s-d routing corridor intersects} \\
% &\qquad\qquad\qquad\qquad\mbox{ square i}|s=(r,\theta)) p(r,\theta) dr d\theta \nonumber \\ 
& \leq \int_0^{c} \frac{2r}{n} dr +\int_{c}^{\sqrt{n}} \left(\frac{2\sqrt{2}\kappa c}{\pi} \frac{\ln n}{\sqrt{n}} + \right. \\
&\qquad\qquad\qquad \left.\frac{(c/\sqrt{2})}{\sqrt{r^2-(c/\sqrt{2})^2}}\frac{(r+\sqrt{n})^2}{\pi n}\right) \frac{2r}{n} dr \\ 
&\leq \frac{c^2}{n} + \frac{2\sqrt{2}\kappa c}{\pi} \frac{\ln n}{\sqrt{n}} + (c/\sqrt{2})\frac{\ln n}{\sqrt{n}} \\ 
&= \left(\frac{c}{\sqrt{n}\ln n} + \frac{2\sqrt{2}\kappa}{\pi} + \frac{1}{\sqrt{2}}\right)\frac{c\ln n}{\sqrt{n}} \\
&\leq \mu \frac{\ln n}{\sqrt{n}} \triangleq p_n
\end{align*}
where $\mu = c (2 + (2\sqrt{2}\kappa)/\pi)$.
\hfill $\square$
\vspace{0.5em}

Returning to the proof of Theorem \ref{thm-maxpath}, since 
$L_i \leq 9\sum_{j=1}^{N_n} A_{ij}$,  we have
\begin{align*}
   \EX[L_i] \leq 9\EX[\sum_{j=1}^{N_n}A_{ij}] 
    = 9 \EX[N_n]\EX[A_{ij}] 
     &= 9\pi n p_{n,i} \leq 9\pi np_n ~. 
\end{align*}
Applying the Chernoff bound \cite[Lemma C3]{duarte-melo},
\begin{align*}
\Pr(L_i > 27\pi np_n) &\leq \exp\left\{-27\pi np_n \ln\frac{3}{e}\right\} ~.
%&= \exp\left(-9\pi np_n\ln\frac{4}{e}
%\right) ~.
\end{align*} 
Substituting the above into (\ref{eq-maxpath}) and choosing $\delta = 27 \pi$ gives
\begin{align*}
\Pr( L(n) \leq \delta \sqrt{n}\ln n) 
%&\geq 1 - \sum_{i=1}^{M_n} \Pr(L_i(n) > \delta \sqrt{n}\ln n)\\
&\geq 1 - \sum_{i=1}^{M_n} \exp\left(-27\pi np_n\ln\frac{3}{e}\right) \\
&= 1 - \exp\left(-27\pi np_n \ln\frac{3}{e} + \ln M_n\right)\\
&= 1 - \exp\left( -27\pi n \frac{\mu \ln n}{\sqrt{n}} \ln\frac{3}{e} + \right.\\
&\qquad \qquad\qquad \qquad \qquad \qquad \left.\ln\frac{\pi n}{c^2}\right) \\
&\rightarrow 1 - 0 = 1 \mbox{  as  } n\rightarrow \infty~,
\end{align*}
which concludes the proof of Theorem \ref{thm-maxpath}.
\hfill $\square$

\section{System Scheduling}
\label{protocol}
In this section we explain a system protocol that achieves a per-node throughput of $\Omega(1/\sqrt{n})$ and complete the proof of Theorem \ref{main}.

%We tessellate the entire network region into squares of side $c$. If a square has multiple nodes in it, we designate one node as the relay node. For every hop of every s-d path, if the node that is to transmit for the hop is not the designated relay node for the square we replace it with the designated relay node. Note that by doing so we increase the maximum intermediate hop length to $(\sqrt{5}+\sqrt{2})c$. However the hop length still remains constant w.r.t. $n$. Since the maximum number of paths that cross a square is $O(\sqrt{n} \ln n)$, every designated relay node will still have to service less than $O(\sqrt{n}\ln n)$ paths.
For every path corresponding to an s-d pair we designate the node on the path that is closest to the source (destination) as the draining (delivery) node. We cycle among  three different categories of time slots: draining, relaying and delivery.  In draining slots, the source transmits its packets to the designated draining nodes. In the relaying slots, the relaying nodes transmit the data towards the destination. Finally in the delivery slots, the delivery nodes transmits the data to the destination.

Theorem \ref{thm-maxpath} shows that the maximum number of s-d paths that a relaying node may have to serve is $O(\sqrt{n} \ln n)$. Since all relaying nodes can transmit at rate $\Omega(1)$ (Corollary \ref{cor-relay}), the relaying node can maintain a throughput of $\Omega(1/\sqrt{n} \ln n)$ per path.

From Corollaries \ref{cor-drain} and \ref{cor-deliver}, it is easy to see that a rate of $\Omega(1/(\sqrt{n}\ln n))$ per path can be maintained in the draining and the delivery phase. 

Thus every s-d pair can achieve a rate of 
\begin{equation}
   \Omega \! \left( \! { 1 \over \sqrt{n} \ln n } \! \right) \mbox{bits/sec/path}~
     \times ~  \Omega(\ln n) \, \mbox{paths} 
      = \Omega \! \left( \! {1 \over \sqrt{n} } \! \right)  \mbox{bits/sec}
\end{equation}
%$\Omega(1/(\sqrt{n}\ln)$ bits/sec/path $\times \Omega(\ln n)$ paths $= \Omega(1/\sqrt{n})$  bits/sec, 
which completes the proof of Theorem \ref{main}.

\bibliographystyle{ieeetrans}

\end{document}